\documentclass[aps,prd,twocolumn,showpacs,amsmath,nofootinbib]{revtex4-1}
\usepackage{amsmath,amssymb,bm}
\usepackage{graphicx}
\usepackage{color}
\usepackage{float}
\usepackage{latexsym}
\usepackage{mathrsfs}
\usepackage{dcolumn}
\usepackage{hyperref}
\usepackage{multirow}
\usepackage[utf8]{inputenc}
\usepackage{microtype}
\usepackage{hyperref}

\begin{document}

\title{Redshift evolution of the Hubble constant: Constraints and new insights from an interacting dark energy model}

\author{Xinyi Dai$^1$}
\author{Yupeng Yang$^1$}\email{corresponding author: ypyang@aliyun.com}
\author{Yicheng Wang$^1$}
\author{Yankun Qu$^1$}\email{quyk@qfnu.edu.cn}
\author{Shuangxi Yi$^1$}\email{yisx2015@qfnu.edu.cn}
\author{Fayin Wang$^2$}\email{fayinwang@nju.edu.cn}

\affiliation{$^1$School of Physics and Physical Engineering, Qufu Normal University, Qufu, Shandong, 273165, China \\
             $^2$School of Astronomy and Space Science, Nanjing University, Nanjing 210023, China}

\begin{abstract}

We develop a modified interacting dark energy (IDE) model to study the redshift evolution of the Hubble constant ($H_0$), in light of the Hubble tension. In this framework, the energy exchange between dark energy and dark matter induces a redshift dependence of $H_0$. We evaluate the model against a comprehensive suite of observations, including baryon acoustic oscillations (BAO) from DESI DR2 and SDSS, cosmic chronometers, type Ia supernovae from the Pantheon sample, and Planck CMB distance priors.
Analysis of late-Universe data yields $\alpha = 0.0107^{+0.0032}_{-0.011}$,  with the best-fit value on the order of $10^{-2}$, revealing a decreasing trend of $H_0$ with redshift. This supports a power-law evolution beyond $\Lambda$CDM. Incorporating CMB data further tightens the constraint to the order of $10^{-5}$, which we attribute to the suppression of dark-sector interactions at high redshifts, a consequence of the strong baryon--photon coupling.
These results indicate that the IDE framework provides a theoretically consistent and observationally viable mechanism for describing the redshift evolution of $H_0$, offering a promising avenue toward alleviating the Hubble tension.

\end{abstract}

\maketitle

%%%%%%%%%%%%%%%%%%%%%%%%%%%%%%%%%%%%%%%%
\section{INTRODUCTION}
\label{sec:1}
%%%%%%%%%%%%%%%%%%%%%%%%%%%%%%%%%%%%%%%%

The standard $\Lambda$CDM model, which has become the prevailing cosmological paradigm, suggests that the Universe is currently undergoing accelerated expansion. This view is strongly supported by diverse observations, including type Ia supernovae (SNIa)~\cite{SupernovaCosmologyProject:1998vns,SupernovaCosmologyProject:1997zqe,SupernovaSearchTeam:1998fmf}, the cosmic microwave background (CMB)~\cite{Caldwell:2003hz,Huang:2005re,Planck:2015mrs}, and baryon acoustic oscillations (BAO)~\cite{BOSS:2016wmc,WiggleZ:2013akc,SDSS:2003eyi}. Despite its success, the model faces several challenges. For instance, the physical origin of dark energy remains unknown, the cosmological constant suffers from fine-tuning and coincidence problems, and the persistent tension in the measurement of the Hubble constant ($H_0$)~\cite{Alestas:2020mvb,Kazantzidis:2019nus,Riess:2021jrx,Kazantzidis:2019nus} remains one of the most significant challenges in modern cosmology.
In particular, $H_0$, which quantifies the current expansion rate of the Universe, is under active debate.
The SH0ES collaboration, using Cepheid-calibrated supernovae via the distance ladder, reports $H_0 = 73.04 \pm 1.04~\mathrm{km\,s^{-1}\,Mpc^{-1}}$~\cite{Riess:2021jrx}, while Planck observations under a flat $\Lambda$CDM model yield $H_0 = 67.4 \pm 0.5~\mathrm{km\,s^{-1}\,Mpc^{-1}}$~\cite{Planck:2018vyg}. This $\sim 5\sigma$ discrepancy has motivated extensive efforts to examine systematics and explore extensions to the standard model~\cite{Yang:2025vnm,Poulin:2018cxd,Hu:2023jqc,Kazantzidis:2019nus,Elizalde:2021kmo,Khurshudyan:2024gpn,CosmoVerseNetwork:2025alb,Yang:2025oax,Yang:2025boq,Wu:2025vfs,Li:2025ops,Li:2025dwz,Du:2025xes,Feng:2025wbz,Guedezounme:2025wav,Yang:2025uyv,Kazantzidis:2019nus,Cai:2021weh,Brito:2024bhh,DiValentino:2019exe}.

Beyond the $H_0$ tension itself, recent studies suggest that the derived Hubble constant may exhibit redshift-dependent behavior, i.e., the inferred value of $H_0$ from observational data (corresponding to different epochs) varies with redshift.
Multiple independent probes, including strong gravitational lensing time delays (SLTD)~\cite{H0LiCOW:2019pvv,Millon:2019slk}, SNIa~\cite{Dainotti:2021pqg,Malekjani:2023ple,Montani:2024ntj}, cosmic chronometers (CC)~\cite{Hu:2022kes}, gamma-ray bursts (GRBs)~\cite{Bargiacchi:2024srw}, and multiwavelength analyses~\cite{Dainotti:2022bzg,Dainotti:2023yrk,Jia:2022ycc,Colgain:2022nlb,Colgain:2022rxy,Jia:2024wix,Krishnan:2020obg}, indicate that the effective Hubble constant decreases with increasing redshift. To describe this effect, a common phenomenological parametrization is adopted:
\begin{equation}
\mathcal{H}_0(z) = H_{0} (1+z)^{-\alpha},
\label{eq:mathcalh0}
\end{equation}
where the constant $\alpha$ characterizes the evolutionary rate and observations typically suggest $\alpha$ is of order $10^{-2}$~\cite{Hu:2022kes,Dainotti:2021pqg,Dainotti:2022bzg,Dainotti:2025qxz,Jia:2024wix}.

Recent BAO measurements from DESI Data Release 1 (DR1) based on the first year of data~\cite{DESI:2024mwx} and Data Release 2 (DR2) spanning three years of observations~\cite{DESI:2025zgx} have provided unprecedented precision in cosmological constraints. The authors of~\cite{Jia:2024wix} applied a nonparametric method to estimate $H_0$ in redshift bins and confirmed a statistically significant ($\sim 6.4\sigma$) decreasing trend of the effective $H_0(z)$, when analyzed within the flat $\Lambda$CDM framework, using DESI DR1, CC, and Pantheon+ SNIa data. This trend bridges local (SH0ES~\cite{Riess:2021jrx}) and early Universe (Planck~\cite{Planck:2018vyg}) determinations of $H_0$, strongly indicating physics beyond the standard model.

Several theoretical approaches have been proposed to account for the apparent redshift dependence of $H_0$. One class involves dynamical dark energy, such as the $w_0w_a$CDM model, which employs the Chevallier-Polarski-Linder (CPL) parametrization to allow the equation of state to evolve with redshift~\cite{Dainotti:2021pqg,Dainotti:2022bzg,Dainotti:2023yrk,Montani:2024ntj,Krishnan:2020vaf}. Another scenario is evolutionary dark energy, where viscous pressure terms in the modified Friedmann equations produce a redshift-dependent effective Hubble constant~\cite{Montani:2024ntj}. Modified gravity theories have also been explored~\cite{Sotiriou:2008rp,Koksbang:2020zej,Sotiriou:2006hs,Najera:2021puo,Paliathanasis:2025xxm,Schiavone:2022wvq}. For example, the authors of~\cite{Dainotti:2022bzg} investigated $f(R)$ gravity with the Hu-Sawicki model in the Jordan frame, finding that the decreasing trend of $H_0$ (with $\alpha \sim 10^{-2}$) persists in SNIa and BAO data. This suggests that more sophisticated extensions of gravity may be required to fully account for the observed evolution of $\mathcal{H}_0(z)$.

Motivated by these observational and theoretical developments, we propose that a self-consistent interacting dark energy (IDE) model, such as that studied in Ref.~\cite{Cai:2004dk}, may also provide an explanation for the observed evolution of the Hubble constant.
Unlike purely phenomenological approaches that enforce a mathematical parametrization on $H_0(z)$ (e.g., the power-law model in Refs.~\cite{Dainotti:2021pqg,Dainotti:2022bzg}), we argue that this evolution arises naturally from a fundamental physical mechanism.
In this framework, the nongravitational energy-momentum exchange between dark energy and dark matter dynamically modifies both the dark energy equation of state and the matter density, thereby affecting the evolution of the $H(z)$.
Crucially, we demonstrate that a specific class of IDE models naturally reproduces the power-law behavior $\mathcal{H}_0(z) \propto (1+z)^{-\alpha}$ without ad hoc assumptions. This transforms $\alpha$ from a mere fitting parameter into a physical quantity derived directly from the dark sector coupling strength.
Using the latest DESI DR2 data, combined with other cosmological probes, we constrain the parameter $\alpha$, which characterizes
the evolutionary rate of $H_0$.

The paper is organized as follows. Section~\ref{sec:2} presents the IDE theoretical framework and demonstrates how the interaction leads to the effective $H_0$ evolution. Section~\ref{sec:3} describes the data and methods, including BAO measurements from DESI and SDSS, CC measurements, Pantheon SNIa, and Planck CMB distance priors. Numerical results are presented in Section~\ref{sec:4}, followed by conclusions in Section~\ref{sec:5}.

%%%%%%%%%%%%%%%%%%%%%%%%%%%%%%%%%%%%%%%%
\section{THE MODEL AND METHODOLOGY}
\label{sec:2}
%%%%%%%%%%%%%%%%%%%%%%%%%%%%%%%%%%%%%%%%
\subsection{Theoretical framework}
We consider a flat Friedmann-Lema\^{i}tre-Robertson-Walker (FLRW) Universe described by the metric~\cite{Deffayet:2000uy}:
\begin{equation}
ds^2 = -dt^2 + a^2(t)\left[dr^2 + r^2(d\theta^2 + \sin^2\theta\, d\phi^2)\right],
\label{eq:metric}
\end{equation}
where $a(t)$ is the cosmic scale factor. We investigate a cosmological model with three components: dark matter (DM), phantom dark energy (DE), and radiation. The Friedmann equations for this model can be written as:
\begin{align}
H^{2} &= \frac{8\pi G}{3}\left(\rho_{m}+\rho_{e} + \rho_{r}\right), \label{eq:friedmann1}\\
2\dot{H} + 3 H^{2} &= -8\pi G\left(p_{m}+p_{e} + p_{r}\right). \label{eq:friedmann2}
\end{align}
Here, $H \equiv \dot{a}/a$ is the Hubble parameter, where $\rho_{e}$ represents the energy density of phantom dark energy with an equation of state $w_{e} \equiv p_{e}/\rho_{e} < -1$. $p_{m}$ and $p_{e}$ denote the pressures associated with DM and DE, respectively. Given that DM is cold, $p_{m} = 0$. $\rho_{r}$ and $p_{r}$ denote the energy density and pressure of radiation, satisfying the equation of state $w_r = p_r/\rho_r = 1/3$.

In IDE models, the total energy-momentum is conserved while individual DM and DE components exchange energy via an interaction term $Q$.
The radiation component is assumed to be noninteracting and evolves independently:
\begin{equation}
\dot{\rho}_r + 4H\rho_r = 0.
\label{eq:rad_int}
\end{equation}
The energy conservation equations for these components can be given by:
\begin{align}
\dot{\rho}_m + 3H\rho_m &= Q, \label{eq:dm_int} \\
\dot{\rho}_e + 3H(1+w_e)\rho_e &= -Q. \label{eq:de_int}
\end{align}

The interaction term $Q$ determines the energy-transfer direction. Specifically, $Q > 0$ denotes energy flow from DE to DM, while $Q < 0$ indicates the reverse. We adopt the standard interaction form for $Q$, as widely employed in the literature
~\cite{Amendola:2006dg, delCampo:2008jx, Bolotin:2013jpa, Wang:2016lxa, Cai:2017yww, Aljaf:2020eqh, Mishra:2023ueo, Pooya:2024wsq, Li:2024qso, Halder:2024aan, Yang:2025ume,Paliathanasis:2026ymi,vanderWesthuizen:2025iam}:
\begin{equation}
Q = \delta H\rho_m.
\end{equation}
The dimensionless coupling parameter $\delta$ preserves the phantom nature of DE ($w_e < -1$). It is subject to observational constraints from both cosmic expansion history and large-scale structure formation.
To facilitate our analysis, we adopt a relation between the energy densities of DE and DM, defined by the ratio $r$ as follows~\cite{Majerotto:2004ji,Cai:2004dk}:
\begin{equation}
r = \frac{\rho_e}{\rho_m} = \frac{\rho_{e,0}}{\rho_{m,0}} \left( \frac{a}{a_0} \right)^{\xi},
\label{eq:energy_ratio}
\end{equation}
where $\xi$ is a constant. Physically, this ratio describes how the energy densities of the dark sector track each other during cosmic evolution.
Setting the current value of the scale factor to unity ($a_0 = 1$), $\rho_{e,0}$ and $\rho_{m,0}$ represent the current dark energy density and dark matter energy density, respectively.

\subsection{Scaling solutions and parametrization }
For the case where the coupling function $\delta$ is constant, we can derive the solutions for these equations (Eqs.~\ref{eq:dm_int} and \ref{eq:de_int}) as follows:
\begin{align}
\rho_m &= \rho_{m,0} a^{-3+\delta}, \label{eq:dm}\\
\rho_e &= \rho_{e,0} a^{-3+\delta+\xi}. \label{eq:de}
\end{align}
The Friedmann equation, which describes the expansion of the Universe, takes the form~\cite{Cai:2004dk}:
\begin{equation}
H(z)^2 = H_0^2 \left( \Omega_{m,0} a^{-3+\delta} + \Omega_{e,0} a^{-3+\delta+\xi} \right). \label{eq:friedmann_equation}
\end{equation}
From the energy conservation equation for dark energy (Eq.~\ref{eq:de_int}), one can derive its equation of state as~\cite{Cai:2004dk}:
\begin{equation}
w_e = -\frac{\delta + \xi}{3} - \frac{\delta}{3}\frac{\Omega_{m,0}}{\Omega_{e,0}}a^{-\xi}.
\label{eq:w_e}
\end{equation}
Thus, the phantom regime ($w_e < -1$) is realized for suitable choices of the parameter set $\{\delta, \xi, \Omega_{m,0} , \Omega_{e,0}\}$~\cite{Cai:2004dk}.

For the specific choice of $\xi = 3$, as discussed in Ref.~\cite{Cai:2004dk}, the DE density follows a scaling behavior parallel to the DM density, with deviations governed by the coupling parameter $\delta$.
This scaling solution allows the model to realize a phantom equation of state ($w_e < -1$) without energy density divergences,
thereby maintaining theoretical consistency.
Furthermore, $\xi = 3$ extends the epoch where DE and DM densities are comparable, naturally alleviating the coincidence problem~\cite{Cai:2004dk} and ensuring a smooth transition from matter domination to late-time accelerated expansion.
Consequently, we fix $\xi = 3$ throughout our analysis.

By substituting $\xi=3$ into Eq.~\eqref{eq:w_e}, we can see that the interaction allows the dark energy equation of state to evolve dynamically. This provides a theoretical basis for an evolving expansion rate that might appear as a redshift-dependent Hubble constant to an observer.
Under this parametrization, the full Friedmann equation takes the form:
\begin{multline}
H(z)^2 = H_0^2 \left[ \Omega_{m,0} (1 + z)^{3-\delta} + \Omega_{e,0} (1 + z)^{-\delta} \right. \\
\left. + \Omega_{r,0} (1 + z)^{4} \right].
\label{eq:modified_friedmann_equation}
\end{multline}
Here, $\Omega_{r,0}$ is the current radiation density parameter. For a flat Universe, the normalization condition requires $\Omega_{m,0} + \Omega_{e,0} + \Omega_{r,0} = 1$. It is important to note that while the radiation term is negligible at low redshifts, it is essential for a consistent analysis of high-redshift data, such as the CMB.

\subsection{Stability analysis and physical interpretation}
To further investigate the global dynamical behavior of this IDE model, we perform a phase-space analysis following Refs.~\cite{Gonzalez-Espinoza:2025bzd,Copeland:2006wr,Boehmer:2014vea}. Within this framework, the equation of state (Eq.~\ref{eq:w_e}) can be decomposed into a baseline term and an interaction-induced correction:
\begin{equation}
w_{e} = -\frac{\delta+3}{3}-\frac{\delta}{3}\frac{\Omega_{m,0}}{\Omega_{e,0}}a^{-3}. \label{eq:w_decomp}
\end{equation}
This decomposition clearly separates the background evolution from the interaction-driven effects, thereby facilitating the phase-space analysis of dark energy dynamics.
In this phase-space description, the evolution of the dark energy density parameter is given by:
\begin{equation}
\frac{d\Omega_{e}}{dN} = \Omega_{e} \left[ \delta - (\delta + 3)\Omega_{e} - \delta \frac{\Omega_{m,0}}{\Omega_{e,0}} a^{-3} (1 - \Omega_{e}) \right]. \label{eq:dOmega}
\end{equation}
The evolution smoothly connects the early matter-dominated point ($\Omega_e = 0$) to the late-time stable attractor ($\Omega_e^* = \delta/(\delta + 3)$), providing a continuous transition from matter domination to phantom accelerated expansion.
Notably, a stability analysis shows that the Jacobian at $\Omega_e^*$ has an eigenvalue $\lambda = -\delta$, implying a stable attractor for $\delta > 0$ and a repeller for $\delta < 0$, thus justifying the physically relevant choice of $\delta > 0$.

In the late Universe where radiation is negligible ($\Omega_{r,0} \approx 0$), the modified Friedmann equation (Eq.~\ref{eq:modified_friedmann_equation}) can be simplified.
By defining the key parameter $\alpha \equiv \delta/2$ and factoring out the common redshift-dependent term $(1+z)^{-\delta}$, we perform an algebraic re-parametrization of the Friedmann equation. Taking the square root, the late-time expansion rate $H(z)$ is given by~\footnote{Note that the radiation term has been 
omitted in the equation presented here for a simple comparison with the standard $\Lambda$CDM model, and this term is fully included in all corresponding numerical 
calculations.
}:
\begin{equation}
\begin{aligned}
H(z) &\simeq H_0 (1+z)^{-\alpha} \sqrt{ \Omega_{m,0} (1+z)^3 + (1 - \Omega_{m,0}) },\\
     &= \mathcal{H}_0(z) \sqrt{ \Omega_{m,0} (1+z)^3 + (1 - \Omega_{m,0}) }.
\end{aligned}
\label{eq:main}
\end{equation}
Here, $H_0$ is the present Hubble constant, and $\alpha$ quantifies the deviation from the standard $\Lambda$CDM cosmology. Crucially, the interaction strength $\delta$ directly generates a redshift-dependent effective Hubble constant $\mathcal{H}_0(z) \equiv H_0 (1+z)^{-\alpha}$.

Equation~(\ref{eq:main}) demonstrates that the IDE model can be effectively described at late times as a noninteracting scenario
in which the core modification is a redshift-dependent Hubble constant $\mathcal{H}_0(z)$.
This functional equivalence is particularly significant when compared to previous phenomenological studies.
For instance, Dainotti et al.~\cite{Dainotti:2021pqg} reported a decreasing trend of $H_0$ with redshift by fitting the standard $\Lambda$CDM model to binned SNIa data.
While such earlier works primarily interpret this evolution
as a symptomatic feature of the Hubble tension or potential systematic effects,
our IDE framework provides a concrete dynamical mechanism: energy exchange between dark matter and dark energy physically modifies the expansion rate, naturally manifesting as the evolving $H_0$ inferred from the data.

Consequently, a nonzero value of $\alpha$ directly implies the existence of dark-sector interaction.
By interpreting the phenomenological $\mathcal{H}_0(z)$ through the lens of these interactions, we bridge empirical observations of $H_0$ evolution with the underlying theoretical framework.
It should be emphasized that while Eq.~\eqref{eq:main} offers this intuitive interpretation for the late Universe,
our numerical analysis employs the complete Friedmann equation (Eq.~\ref{eq:modified_friedmann_equation}),
including the radiation term, to ensure accuracy across all redshift scales.

%%%%%%%%%%%%%%%%%%%%%%%%%%%%%%%%%%%%%%%%
\section{DATASETS AND CONSTRAINTS}
\label{sec:3}
%%%%%%%%%%%%%%%%%%%%%%%%%%%%%%%%%%%%%%%%

\subsection{Baryon acoustic oscillation}

To constrain the cosmic expansion history, we include BAO measurements from the latest DESI observations and earlier SDSS surveys.
\begin{itemize}
\item \textbf{DESI} We adopt the latest dataset from DESI DR2~\cite{DESI:2025zgx}, which includes 25 data points spanning the redshift range $0.1 < z < 4.2$. These measurements are obtained from multiple subsamples: the Bright Galaxy Sample (BGS, $0.1 < z < 0.4$), Luminous Red Galaxies (LRG, $0.4 < z < 1.1$), Emission Line Galaxies (ELG, $0.8 < z < 1.6$), Quasars (QSO, $0.8 < z < 2.1$), and the Lyman-$\alpha$ forest (Ly$\alpha$, $1.8 < z < 4.2$)~\cite{Hahn:2022dnf,Raichoor:2022jab,Chaussidon:2022pqg}. We use the distance ratios $D_M/r_d$, $D_H/r_d$, and $D_V/r_d$, where all distances are scaled by the sound horizon $r_d$. Following Refs.~\cite{DESI:2024mwx,DESI:2025zgx}, we treat $r_d$ as a free parameter and refer to this dataset as ``DESI'' throughout.

\item \textbf{SDSS} We also include pre-DESI BAO measurements~\cite{2012JCAP...03..027G} from earlier surveys, including the 6dF Galaxy Survey, the SDSS, BOSS CMASS, and WiggleZ~\cite{2011MNRAS.418.1707B,2010MNRAS.401.2148P,2011MNRAS.416.3017B,2011ApJS..192...14J,SDSS:2005xqv}. This dataset provides six data points expressed as $d_A(z_*)/D_V(z_{\text{BAO}})$, where $z_* \approx 1091$ denotes the redshift of photon decoupling, and $d_A(z_*)$ is the comoving angular diameter distance at that epoch. We refer to this dataset as ``SDSS'' hereafter.
\end{itemize}

\subsection{Cosmic chronometer}
The CC method provides a model-agnostic estimate of the Hubble parameter $H(z)$ by measuring the differential age evolution of passively evolving galaxies. This technique~\cite{Jimenez:2001gg} exploits the relation:
\begin{equation}
H(z) = -\frac{1}{1+z} \frac{\Delta z}{\Delta t},
\end{equation}
where $\Delta z$ and $\Delta t$ are the redshift and age differences between galaxies at distinct epochs.

For this analysis, we adopt a curated subset of 32 $H(z)$ measurements~\cite{Moresco:2022phi}, spanning $0.07 < z < 1.965$. The dataset is selected to minimize systematic biases from stellar population modeling, as discussed in Ref.~\cite{Moresco:2022phi}.
This approach enables independent constraints on the expansion history without relying on external distance anchors, thus serving as a complementary probe to SNIa or BAO.

\subsection{Type Ia supernova}
Our analysis uses the Pantheon Sample~\cite{Pan-STARRS1:2017jku}, a widely recognized compilation of 1048 spectroscopically confirmed SNIa measurements covering a broad redshift range ($0.01 \leq z \leq 2.26$).
Thanks to their standardized candle nature, SNIa allow us to determine luminosity distances ($d_L$) via the distance modulus relation,
\begin{equation}
\mu = 5 \log_{10}d_L + 25,
\end{equation}
following the methodology in~\cite{Yang:2025boq}.
The Pantheon dataset includes contributions from multiple surveys, such as the PanSTARRS1 Medium Deep Survey, SDSS, the Supernova Legacy Survey (SNLS), and various low-redshift and Hubble Space Telescope (HST) samples~\cite{Chang:2019utc}.

\subsection{Cosmic microwave background}

To maintain computational efficiency while retaining the constraining power of the CMB data, we adopt the distance priors derived from the final Planck 2018 release~\cite{Chen:2018dbv,Planck:2018vyg}. The full Planck data can be compressed into three key parameters: the shift parameter $R$, the acoustic scale $l_A$, and the physical baryon density parameter $\omega_b = \Omega_b h^2$.

The shift parameter $R$ and the acoustic scale $l_A$ are defined as~\cite{2020A&A...641A...6P,Xu:2016grp,Feng_2011}:
\begin{equation}
R = \frac{1 + z_\star}{c} D_{\rm A}(z_\star) \sqrt{\Omega_m H_0^2},
\end{equation}
\begin{equation}
l_{\rm A} = (1 + z_\star) \frac{\pi D_{\rm A}(z_\star)}{r_s(z_\star)},
\end{equation}
where $z_\star$ is the redshift at the epoch of photon decoupling. We adopt the fitting formula proposed by~\cite{Hu:1995en} for $z_\star$ 
\footnote{The corrected form can be found in Ref.~\cite{Aizpuru:2021vhd}.}:
\begin{equation}
z_\star = 1048 \left[ 1 + 0.00124(\Omega_b h^2)^{-0.738} \right] \left[ 1 + g_1 (\Omega_m h^2)^{g_2} \right],
\label{eq:zstar}
\end{equation}
where the coefficients $g_1$ and $g_2$ are given by:
\begin{equation}
g_1 = \frac{0.0738 (\Omega_b h^2)^{-0.238}}{1 + 39.5(\Omega_b h^2)^{0.763}}, \quad g_2 = \frac{0.560}{1 + 21.1(\Omega_b h^2)^{1.81}} .
\label{eq:g1g2}
\end{equation}
The comoving sound horizon $r_s(z)$ is computed using the modified expansion rate $H(a)$ derived from our IDE framework:
\begin{equation}
r_s(z) = \frac{c}{H_0} \int_0^{1/(1+z)} \frac{da}{a^2 H(a)\sqrt{3\left(1 + \frac{3\Omega_b h^2}{4\Omega_\gamma h^2}a\right)}},
\label{eq:rs}
\end{equation}
where $(\Omega_\gamma h^2)^{-1} = 42000(T_{\rm CMB}/2.7\, \mathrm{K})^{-4}$ and $T_{\rm CMB} = 2.7255 \text{K}$.

Note that this approach primarily captures the background geometric information and does not fully account for secondary effects such as CMB lensing signatures~\cite{Planck:2018lbu, Lewis:2006fu} or the integrated Sachs-Wolfe (ISW) effect~\cite{Planck:2015fcm, PhysRevLett.76.575}.
However, these effects are known to exert a subdominant impact on the total likelihood compared to the primary acoustic peaks for the class of models studied here.
Indeed, it has been established that CMB distance priors serve as a consistent and robust surrogate for the full CMB likelihood across a wide range of dark energy models~\cite{Wang:2007mza, Zhai:2019nad}.
For dynamical models characterizing dark energy dynamics, these priors have been demonstrated to be an accurate representation of the full dataset~\cite{Zhai:2019nad}.
As our model asymptotically converges to the standard matter-dominated behavior at high redshifts, the geometric information provided by $(R, l_A, \omega_b)$ yields a high-fidelity representation of the Planck data, ensuring that any potential information loss is statistically marginal for the determination of the model parameters.
% -------------------------

We denote the parameter vector as $\mathbf{x} = (R, l_A, \omega_b)$. The corresponding $\chi^2$ function for the CMB data is expressed as:
\begin{equation}
\chi^2_{\text{CMB}} = \Delta \mathbf{x} \cdot C_{\text{CMB}}^{-1} \cdot \Delta \mathbf{x}^T,
\end{equation}
where $\Delta \mathbf{x} = \mathbf{x} - \mathbf{x}^{\text{Pl}}$ is the difference vector between the model parameters $\mathbf{x}$ and the observed Planck values $\mathbf{x}^{\text{Pl}}$:
\begin{equation}
\mathbf{x}^{\text{Pl}} = (1.7493 \pm 0.0046,\, 301.462 \pm 0.090,\, 0.02239 \pm 0.00015).
\end{equation}
The covariance matrix $C_{\text{CMB}}$ is taken from Ref.~\cite{Chen:2018dbv}.

%%%%%%%%%%%%%%%%%%%%%%%%%%%%%%%%%%%%%%%%
\section{RESULTS AND DISCUSSION}
\label{sec:4}
%%%%%%%%%%%%%%%%%%%%%%%%%%%%%%%%%%%%%%%%

%****++++++++++++++++++++++++ Table 1 ********************************************
\begin{table*}[htb]
  \centering
  
  \renewcommand{\arraystretch}{1.5} % \D4\F6\BC\D3\D0и\DF
  \caption{Constraints on the parameter sets \{$\alpha$, $\Omega_{m,0}$, $H_0$, $\Omega_{b}h^{2}$\} (with 1$\sigma$) utilizing the combined datasets:
  DESI+SDSS+CC+SNIa and DESI+SDSS+CC+SNIa+CMB. Here, $H_0$ is given in units of $\rm km~s^{-1}~Mpc^{-1}$.}

    \begin{tabular}{l|cccc}
  \hline
  \hline
  Dataset & $\alpha$ & $\Omega_{m,0}$ & $H_0$ &$\Omega_{b}h^{2}$  \\
  \hline
  DESI+SDSS+CC+SNIa & $ 0.0107^{+0.0032}_{-0.011}$ &$0.2952^{+0.0049}_{-0.0087}$  &$69.85\pm{0.15}$ &$0.0224^{+0.0010}_{-0.00056}$ \\
  \hline
  DESI+SDSS+CC+SNIa+CMB & $0.000056^{+0.000016}_{-0.000056}$ &$0.2869\pm{0.0017}$  &$69.67\pm{0.13}$ &$0.02297\pm{0.0001}$ \\
  \hline
  \hline
    \end{tabular}%
  \label{tab:1}%
\end{table*}

In this study, we performed a Markov chain Monte Carlo (MCMC) analysis of the IDE model using the \texttt{EMCEE} package~\cite{emcee}. Uniform priors are adopted for the four free parameters within the following conservative ranges: $\alpha \in (0, 1)$, $\Omega_{m} \in (0, 1)$, $H_0 \in (50, 100) \, \mathrm{km \, s^{-1} \, Mpc^{-1}}$, and $\Omega_b h^2 \in (0.001, 0.1)$. To ensure reproducibility and reliable posterior sampling, we performed a systematic convergence assessment of the MCMC chains. We utilized 128 walkers, each evolving for 5000 steps. Convergence was monitored via the integrated autocorrelation time, $\tau$, estimated using the \texttt{get\_autocorr\_time} utility in \texttt{emcee}. Since $\tau$ provides a measure of sample independence, the total chain length was ensured to be many times larger than $\tau$. Our diagnostics show $\tau \approx 50$ steps for all parameters. Consequently, we discarded the initial one hundred steps as a burn-in phase to fully remove initial transients and ensure the remaining samples represent the stationary posterior distribution.
The sampling efficiency was further validated by the mean acceptance fraction, which was found to be 0.4394. This value falls well within the optimal range of 0.20--0.50 recommended for affine-invariant ensemble samplers~\cite{2013PASP..125..306F}, indicating efficient exploration of the parameter space.

%%%%%%%%%%%%%%%%%  firgure  %%%%%%%%%%%%%%%%%%%%

\begin{figure}[htbp]
    \centering
    \includegraphics[width=\columnwidth]{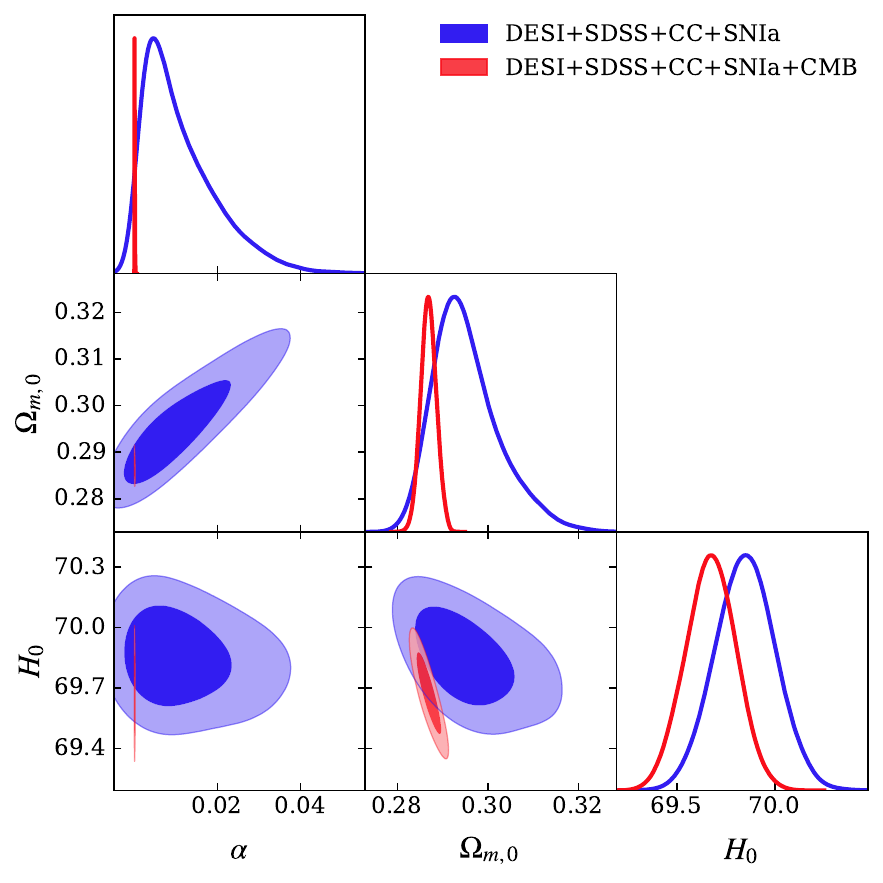}
    \caption{2-D posterior distributions and 1-D marginalized constraints for the key parameters $\{\alpha$, $\Omega_{m,0}$, $H_0\}$ at the 68\% and 95\% confidence levels, derived using the combined datasets DESI+SDSS+CC+SNIa (blue) and DESI+SDSS+CC+SNIa+CMB (red). Here, $H_0$ is given in units of $\rm km~s^{-1}~Mpc^{-1}$.
   }
\label{fig:1}
\end{figure}

The best-fit values of the model parameters were obtained by maximizing the likelihood, $L \propto e^{-\chi^{2}/2}$, over the full sample of converged points. The posterior distributions and corresponding confidence regions were then computed and visualized using the \texttt{GetDist} package~\cite{getdist}.
To constrain the IDE model, we incorporate five key observational datasets: BAO measurements from DESI DR2 and SDSS, along with CC, SNIa, and CMB data. The total chi-squared function used in the joint likelihood analysis is given by
\begin{equation}
\chi^{2}_{\mathrm{total}} = \chi^{2}_{\mathrm{DESI}} + \chi^{2}_{\mathrm{SDSS}} + \chi^{2}_{\mathrm{CC}} + \chi^{2}_{\mathrm{SNIa}} + \chi^{2}_{\mathrm{CMB}},
\end{equation}
where $\chi^2_{\mathrm{total}}$ denotes the sum of the chi-squared contributions from all datasets. The resulting parameter constraints are summarized in Table~\ref{tab:1}, and Fig.~\ref{fig:1} shows the corresponding $1\sigma$ and $2\sigma$ confidence contours.

To evaluate the fit quality and model preference, we calculate the reduced chi-square, $\chi^2_{\mathrm{red}} = \chi^2_{\mathrm{min}} / (N - k)$, and the Akaike information criterion (AIC)~\cite{Akaike1974}, defined as $\mathrm{AIC} = \chi^2_{\mathrm{min}} + 2k$.
As summarized in Table~\ref{tab:2}, the goodness of fit is confirmed by $\chi^2_{\mathrm{red}} \approx 1$ for both models across all datasets ($N=1111$ for late Universe probes; $N=1114$ including CMB).
These results indicate that the IDE framework provides a statistically consistent description of the data without excessive complexity, thereby validating our numerical procedure.

The statistical necessity of the additional parameter $\alpha$ is evaluated using the AIC difference, $\Delta \mathrm{AIC} = \mathrm{AIC}_{\mathrm{IDE}} - \mathrm{AIC}_{\Lambda\mathrm{CDM}}$. For the baseline DESI+SDSS+CC+SNIa dataset, the IDE model yields $\Delta \mathrm{AIC} = -2.55$, a preference that strengthens to $-3.95$ upon the inclusion of CMB data. According to standard model selection criteria\footnotemark, these negative values indicate substantial support for the interaction framework, confirming that the $\alpha$ parametrization is statistically justified by current observations.

%****++++++++++++++++++++++++ Table 2 ********************************************
\begin{table*}[htb]
  \centering
  \renewcommand{\arraystretch}{1.5}
  \caption{Statistical assessment of model performance: the minimum chi-square ($\chi^2_{\mathrm{min}}$), reduced chi-square ($\chi^2_{\mathrm{red}}$), and the Akaike Information Criterion (AIC). The relative statistical preference for the IDE framework is quantified by $\Delta \mathrm{AIC} = \mathrm{AIC}_{\mathrm{IDE}} - \mathrm{AIC}_{\Lambda\mathrm{CDM}}$.}
  \label{tab:2}
  \begin{tabular}{l|ccccc}
  \hline\hline
  Dataset & Model & $\chi^2_{\mathrm{min}}$ & $\chi^2_{\mathrm{red}}$ & AIC & $\Delta \mathrm{AIC}$ \\
  \hline
  \multirow{2}{*}{DESI+SDSS+CC+SNIa} & $\Lambda$CDM & 1117.97 & 1.0090 & 1123.97 & \textemdash \\
                                     & IDE          & 1113.42 & 1.0058 & 1121.42 & $-2.55$ \\
  \hline
  \multirow{2}{*}{DESI+SDSS+CC+SNIa+CMB} & $\Lambda$CDM & 1141.57 & 1.0275 & 1147.57 & \textemdash \\
                                         & IDE          & 1135.62 & 1.0231 & 1143.62 & $-3.95$ \\
  \hline\hline
  \end{tabular}
\end{table*}

\footnotetext[1]{Following the Burnham and Anderson information-theoretic criteria \cite{Burnham2002}, the relative strength of evidence for a model is assessed via $\Delta \mathrm{AIC}$. A difference of $0 < |\Delta \mathrm{AIC}| < 2$ indicates ``weak evidence'' or that the models are nearly indistinguishable; $2 < |\Delta \mathrm{AIC}| < 4$ signifies ``substantial evidence'' in favor of the model with the lower AIC; $4 < |\Delta \mathrm{AIC}| < 7$ suggests ``strong evidence''; while $|\Delta \mathrm{AIC}| > 10$ indicates that the model with the higher AIC has ``essentially no support'' from the data.}

The constraint analysis based on late Universe observational data (DESI+SDSS+CC+SNIa) indicates that the interaction parameter is constrained to $\alpha = 0.0107^{+0.0032}_{-0.011}$. This result suggests a potential redshift-dependent evolution of the Hubble constant $H_0$. Furthermore, by comparing our late-Universe dataset with other relevant studies, we find that the evolutionary parameter $\alpha$ is consistently of the order of $10^{-2}$. The following presents a detailed discussion of the underlying explanations for these constraints:
\begin{itemize}
    \item The binned analysis of the Pantheon sample conducted in Ref.~\cite{Dainotti:2021pqg} introduced the power-law profile $\mathcal{H}_0(z) = H_0(1 + z)^{-\alpha}$ and reported consistent estimates of the evolutionary parameter: $\alpha=0.009\pm0.004$ for 3 bins, $\alpha=0.008\pm0.006$ for 4 bins, $\alpha=0.014\pm0.010$ for 20 bins, and $\alpha=0.016\pm0.009$ for 40 bins (referred to as the PL model). These results consistently cluster around the $10^{-2}$ level, indicating a potential redshift evolution of $\mathcal{H}_0(z)$ beyond what can be explained by statistical fluctuations alone.

    \item Building on Ref.~\cite{Dainotti:2021pqg}, Dainotti et al.~\cite{Dainotti:2022bzg} performed a joint binned analysis of the Pantheon sample combined with BAO data, obtaining $\eta=0.008\pm0.006$ under the PL model. In the context of the Hu-Sawicki $f(R)$ model, the parameter $\eta$ shows stability with respect to the $\Lambda$CDM prediction, suggesting that the redshift evolution signal persists even under modified gravity. In scalar-tensor theories, theoretical predictions of $\eta \sim 10^{-2}$ are also consistent with these observational constraints, although the precise functional form of the scalar potential requires further validation.

    \item Dainotti et al.~\cite{Dainotti:2025qxz} constructed a new SNIa dataset, referred to as the master binned SNIa sample, consisting of 20 equally populated redshift bins covering the range $z = 0.00122$ to $z = 2.3$. By performing a PL model fit to this dataset, the authors derived best-fit values for the interaction parameter $\alpha = 0.010$ and the Hubble constant $H_0 = 69.869~\mathrm{km\,s^{-1}\,Mpc^{-1}}$.
\end{itemize}

From these results, it is evident that when employing late Universe observations to constrain the model, the derived values of $\alpha$ are in good agreement with our analysis. While slight numerical differences may arise due to variations in data quality and methodology, the overall trend and magnitude of $\alpha$ remain consistent across studies. This consistency further supports the existence of a redshift-dependent effect in the cosmic expansion history.
According to Eq.~(\ref{eq:mathcalh0}), extrapolating this late Universe trend to the last scattering surface ($z = 1100$) yields $\mathcal{H}_0(z = 1100) = 64.81 \pm 0.14~\mathrm{km\,s^{-1}\,Mpc^{-1}}$.
Remarkably, this extrapolated value provides a consistent bridge to our joint late and early Universe analysis, indicating that the late-time evolution, encoded by a nonzero $\alpha$ naturally points toward the low $H_0$ environment required by the CMB data. This behavior highlights a predictive internal consistency of the interacting dark energy framework across different cosmic epochs.

To assess the robustness of our results, we performed a sensitivity analysis by varying the fixed parameter $\xi$ over the values 2.90, 2.95, 3.00, 3.05, and 3.10 
using the combined dataset DESI+SDSS+CC+SNIa. The corresponding best-fit values and $1\sigma$ uncertainties for the model parameters are summarized in Table~\ref{tab:robustness}. Our analysis reveals that the mean value of the interaction parameter $\alpha$ exhibits a noticeable variation, increasing slightly with larger $\xi$. 
It should be noted that in order to make the final results more pronounced, we have chosen a relatively large step size for varying $\xi$.

%%%%%%%%%%%%%%%%%%%%%%%% table III  %%%%%%%%%%%%%%%%%%
\begin{table}[htb]
\centering
\renewcommand{\arraystretch}{1.5}
\caption{Model parameter variations for fixed values of $\xi =$ 2.90, 2.95, 3.00, 3.05, and 3.10 from the DESI+SDSS+CC+SNIa dataset. 
Here, $H_0$ is given in units of $\rm km~s^{-1}~Mpc^{-1}$.
}
\label{tab:robustness}
\begin{tabular}{c|cccc}
\hline\hline
$\xi$ & $\alpha$  & $\Omega_{m,0}$ & $H_0$ &$\Omega_{b}h^{2}$  \\
\hline

2.90 & $0.0085^{+0.0032}_{-0.0080}$ & $0.2890^{+0.0043}_{-0.0072}$ & $69.71\pm{0.15}$ & $0.02350^{+0.00090}_{-0.00052}$ \\ \hline
2.95 & $0.0098^{+0.0038}_{-0.0097}$ & $0.2918^{+0.0048}_{-0.0084}$ & $69.80\pm{0.15}$ & $0.0230^{+0.0010}_{-0.00055}$ \\ \hline
3.00 & $0.0107^{+0.0032}_{-0.011}$  & $0.2952^{+0.0049}_{-0.0087}$ & $69.85\pm{0.15}$ & $0.0224^{+0.0010}_{-0.00056}$ \\ \hline
3.05 & $0.0131^{+0.0054}_{-0.012}$  & $0.2893^{+0.0055}_{-0.0096}$ & $69.95\pm{0.16}$ & $0.0219^{+0.0011}_{-0.00063}$ \\ \hline
3.10 & $0.0147^{+0.0063}_{-0.013}$  & $0.3005^{+0.0063}_{-0.010 }$ & $70.09\pm{0.16}$ & $0.0214^{+0.0011}_{-0.00071}$ \\ 
\hline\hline
\end{tabular}
\end{table}

However, the incorporation of early Universe CMB data significantly tightens the constraint on the interaction parameter to $\alpha \sim \mathcal{O}(10^{-5})$, as shown in Table~\ref{tab:1}. This stringent constraint reflects the high precision of the CMB in determining the angular sound horizon. While we have verified that the interaction term is subdominant during the radiation-dominated era, the exceptional sensitivity of the acoustic peak positions to the expansion history ensures that even a minor modification to $H(z)$ would be detectable~\cite{2020A&A...641A...6P,Hu:2001bc}.

Mathematically, the validity of using CMB distance prior compressions is supported by the structure of Eq.~\eqref{eq:modified_friedmann_equation}. As discussed in the methodology section, the radiation density $\rho_r \propto (1+z)^4$ dominates other components at high redshifts. Since its power index is higher than that of the modified matter density ($3-2\alpha$), the expansion rate $H(z)$ asymptotically converges to the standard $\Lambda$CDM evolution. This ensures that the early Universe physics remains standard, providing a self-consistent justification for the distance prior method as a robust and unbiased surrogate for the full CMB likelihood.

In our joint analysis (DESI+SDSS+CC+SNIa+CMB), the Hubble constant is measured as $H_0 = 69.67 \pm 0.13~\mathrm{km\,s^{-1}\,Mpc^{-1}}$. This value represents an optimal balance between late Universe expansion trends and the stringent geometric constraints from the CMB.
By incorporating a dynamical dark-sector interaction, the IDE framework allows a suppressed early Universe coupling $\alpha$ while sustaining a relatively high present-day Hubble constant. This mechanism reconciles the differing $H_0$ requirements of the early and late Universe within a single self-consistent global fit.
The derived baryon density, $\Omega_b h^2 = 0.02297 \pm 0.00010$, is in excellent agreement with big bang nucleosynthesis predictions. These results collectively demonstrate the internal consistency of the IDE framework across a wide range of cosmic epochs.

%%%%%%%%%%%%%%%%%%%%%%%%%%%%%%%%%%%%%%%%
\section{CONCLUSION}
\label{sec:5}

%%%%%%%%%%%%%%%%%%%%%%%%%%%%%%%%%%%%%%%%
In this study, we have constructed and analyzed an IDE model to systematically explore the redshift evolution of the Hubble constant $H_0$ and its cosmological implications. By combining the latest BAO measurements from DESI DR2 with SDSS, CC, SNIa, and CMB data, we refined the characterization of the IDE parameters. Specifically, late-time observations constrain the interaction parameter $\alpha$ to the order of $10^{-2}$. This finding strongly supports the PL evolution model $\mathcal{H}_0(z) = H_0(1 + z)^{-\alpha}$, and importantly, the IDE model provides a theoretical foundation for this power-law form of $\mathcal{H}_0(z)$. Crucially, our framework reveals that this parametrization is fundamentally rooted in the dark-sector interaction $\delta$ through the simple mapping $\alpha = \delta/2$, providing a dynamical origin for the observed expansion history.
The obtained constraints on $\alpha$ reveal a significant trend in the redshift evolution of $H_0$, indicating the potential presence of new physics beyond the standard $\Lambda$CDM model that could help resolve the Hubble tension.

Furthermore, the incorporation of early Universe CMB data significantly tightens the constraints on $\alpha$ to the order of $10^{-5}$, highlighting the critical role of precise high-redshift observations in shaping our understanding of dark-sector interactions. This result indicates that any energy exchange between DM and DE is strongly constrained to be suppressed in the early Universe, consistent with the tight baryon-photon coupling at $z \simeq 1100$.
Specifically, this suggests that while dark-sector interactions may be active and relevant at late times, the early Universe remains remarkably consistent with the standard $\Lambda$CDM paradigm.
This underscores the necessity of combining cosmological probes across different redshift regimes, and future research should extend this approach by incorporating a broader set of observations, such as the Tip of the Red Giant Branch (TRGB)~\cite{Freedman:2024eph}, fast radio bursts (FRBs)~\cite{Zhang:2022uzl,Bochenek:2020zxn}, and gamma-ray bursts (GRBs)~\cite{Dainotti:2020azn,Ryan:2019fhz}.

In summary, the IDE model proposed here not only offers a plausible explanatory framework for late-Universe dynamics but also provides a self-consistent theoretical basis for the redshift evolution of $H_0$. It supports the concept of an interacting dark energy component and offers a possible pathway toward alleviating the Hubble tension by reconciling constraints from different redshift regimes.

\section*{Acknowledgements}
Y. Yang thanks Lei Feng for helpful comments.
This work is supported by the Shandong Provincial Natural Science Foundation (Grant Nos. ZR2025MS16, ZR2025MS47) 
and the China Postdoctoral Science Foundation (Grant No. 2025M783226).

\bibliography{refs1111}

\end{document}